\documentclass[11pt]{article}
\usepackage{graphicx}
\usepackage{mathtools}
\usepackage{dsfont}
\usepackage{graphics,amsmath,amssymb,amsthm,accents,amsfonts}
\usepackage{epic}
\usepackage{amsmath}
\usepackage{color}
\usepackage{cite}
\makeatletter

\newcommand{\Rmnum}[1]{\expandafter\@slowromancap\romannumeral #1@}
\makeatother
\def \h#1{\widehat{#1}}
\def \t#1{\widetilde{#1}}

\def \th#1{\widehat{\widetilde{#1}}}

\def\dwh{\underaccent{{\cc@style\widehat{\mskip10mu}}}}
\def \dt#1{\underaccent{\tilde}{#1}}

\newcommand{\bom}{\boldsymbol{\omega}}
\newcommand{\balpha}{\boldsymbol{\alpha}}

\newcommand{\bft}{\mathbf{t}}
\newcommand{\bthe}{\boldsymbol\theta}

\makeatother \numberwithin{equation}{section}
\newtheorem{prop}{Proposition}
\allowdisplaybreaks

\begin{document}

%----------------------------------------------------------------------------------------
%	TITLE PAGE
%----------------------------------------------------------------------------------------
%\begin{CJK*}{GBK}{kai}

\title {Spectrum transformation and conservation laws of the lattice potential KdV equation}
\author{Senyue Lou$^1$, ~Ying Shi$^2$,~ Da-jun Zhang$^3$\footnote{Corresponding author. Email: djzhang@staff.shu.edu.cn}\\
{\small \it ${}^{1}$Faculty of Science, Ningbo University, Ningbo 315211, China}\\
{\small \it  ${}^{2}$School of Science, Zhejiang University of Science and Technology, Hangzhou 310023, P.R. China}\\
{\small \it ${}^{3}$Department of Mathematics, Shanghai University, Shanghai 200444, P.R. China}}
\date{\today}

%\begin{document}
\maketitle
%--------------------------------------------------------------------------------------------------------------------------------------------------------
%--------------------------------------------------------------------------------------------------------------------------------------------------------
%--------------------------------------------------------------------------------------------------------------------------------------------------------

\begin{abstract}

Many multi-dimensional consistent discrete systems have soliton solutions with nonzero backgrounds,
which brings difficulty in the investigation of integrable characteristics.
In this letter we derive infinitely many conserved quantities for the lattice potential Korteweg-de Vries equation.
The derivation is based on the fact that the scattering data $a(z)$ is independent of discrete space and time
and the analytic property of Jost solutions of the discrete Schr\"odinger spectral problem.
The obtained conserved densities are different from those in the known literatures.
They are asymptotic to zero when $|n|$ (or $|m|$) tends to infinity.
To obtain these results, we reconstruct a discrete Riccati equation by using a conformal map
which transforms the upper complex plane to the inside of unit circle.
Series solution to the Riccati equation is constructed based on the analytic and asymptotic properties of Jost solutions.


\end{abstract}

\section{Introduction}

For a nonlinear partial differential equation, possessing infinitely many conservation laws
indicates the equation is integrable.
In the fully discrete case, multi-dimensional consistency\cite{NW-GMJ-2001,Nijhoff-PLA-2002,ABS-CMP-2003} provides a kind of
strong integrability, under which and extra conditions
lattice equations defined on quadrilateral were classified\cite{ABS-CMP-2003} and
discrete Boussinesq(DBSQ)-type equations were found\cite{Hietarinta-JPA-2011} as multi-component models.
For these multi-dimensional consistent equations, infinitely many conservation laws were already constructed
in several different ways\cite{RH-SIGMA-2005,RH-JPA-2007,RS-JPA-2009,Rasin-JPA-2010,Xenitidis-JPA-2011,MWX-TMP-2011,
Xenitidis-PLA-2012,ZCS-JPA-2013,CZ-FMC-2013,HY-arxiv-2013}.
However, these known conservation laws are only formal ones
--- the nonzero boundary condition at infinity (asymptotic behavior) of the solutions  of the considered lattice equations was neglected.

How does asymptotic behavior play roles?
A discrete conservation law is defined in the form
\begin{equation}
(E_m-1)F_{n,m}=(E_n-1)J_{n,m},
\label{cls-1}
\end{equation}
where $E_m$ and $E_n$ are shift operators defined as $E_m f_{n,m}=f_{n,m+1}$ and  $E_n f_{n,m}=f_{n+1,m}$.
We request $F_{n,m}$ and $J_{n,m}$ tend to zero fast enough when $|n|\to \infty$
so that $\sum_{n=-\infty}^{\infty}[(E_n-1)J_{n,m}]=0$ and the sumation   $\sum_{n=-\infty}^{\infty}F_{n,m}$ makes sense.
As a result, $\sum_{n=-\infty}^{\infty}F_{n,m}$ is indenpendent of $m$ and represents a conserved quantity with respect to $m$.

To explain more,
% (to see the drawbacks??),
as an example,
let us look at the lattice potential Korteweg-de Vries (lpKdV) equation\cite{NQC-PLA-1983,NC-AAM-1995},
also known as H1 equation in the Adler-Bobenko-Suris (ABS) list\cite{ABS-CMP-2003},
\begin{equation}\label{H1}
(\th u-u)(\t u-\h u)=p^2-q^2~,
\end{equation}
where $u=u_{n,m}$, $\t u=u_{n+1,m}$, $\h u=u_{n,m+1}$, $\th u=u_{n+1,m+1}$, and lattice parameters $p$ for direction $n$ and $q$
for direction $m$ are arbitrary constants.
Unlike most of  continuous integrable systems,
$u=0$ is not a solution to the lpKdV equation \eqref{H1}.
It has a simplest solution
\begin{equation}
u=pn+qm+c
\label{0SS}
\end{equation}
with a constant $c$. The $N$-soliton solution has the form\cite{NAH-JPA-2009,HZ-JPA-2009}
\begin{equation}
u=pn+qm+c+w
\label{NSS}
\end{equation}
where
\begin{equation}
w \to 0,~~|n|\to \infty~ (\mathrm{or}~|m|\to \infty).
\end{equation}
Solution \eqref{0SS} is usually called a \textit{background} of solitons.
It has been known that most of members in the ABS list and DBSQ-type equations
have either linear or exponential backgrounds (cf. \cite{NAH-JPA-2009,HZ-JPA-2009,HZ-JMP-2010,HZ-SIGMA-2011,ZZN-SAM-2012,BJ-IP-2010,Butler-Non-2012}).
The nonzero background will bring difficulties in finding convergent conserved quantities.
For the lpKdV equation \eqref{H1}, due to the background \eqref{0SS},
the equation itself is not in a conserved form and the discrete integral of function of $u$,
for example, $\sum^{\infty}_{n=-\infty}u_{n,m}$, does not make sense
because with the background \eqref{0SS} the summation is not convergent at all.
With regard to discrete conservation laws,
as an example, let us take a look at the second conservation law of the lpKdV equation derived from its Lax pair (cf. \cite{ZCS-JPA-2013}),
which is in the form \eqref{cls-1} with
\begin{equation}
F_{n,m}=\frac{1}{(u-\t{\t{u}})(\t u-\t{\t{\t{u}}})},~~ J_{n,m}=\frac{1}{(\t u-\h u)(u-\t{\t{u}})}.
\label{2nd-FJ}
\end{equation}
Obviously, for $u$ defined in \eqref{NSS} with the background \eqref{0SS}, neither $F_{n,m}$ nor $J_{n,m}$
is asymptotic to zero when $|n|\to 0$ or $|m| \to 0$.
Although one can artificially add constants $-1/4p^2$ and $1/2p(p-q)$ to the above $F_{n,m}$ and $J_{n,m}$ respectively
so that they get zero asymptotic behaviors,
these constants can not be derived naturally.
This is the drawback.
% of the  conservation law.

In this letter, for the lpKdV equation \eqref{H1}, we derive its infinitely many conserved quantities
of which each corresponding conserved density $F_{n,m}\to 0$ when $|n|\to \infty$ and the conserved quantity
$\sum^{\infty}_{n=-\infty}F_{n,m}$ is convergent.
Our approach is a by-product of the inverse scattering transform (IST).
In the IST, the scattering data $a(\lambda)$ (connected with the transmission coefficient $T$ by $T=1/a(\lambda)$)
is time-independent and it is asymptotically related to wave functions (Jost solutions) in a certain way.
$a(\lambda)$ can be expanded in the complex plane and coefficient of each term provides a conserved quantity (functional).
Such an approach was first proposed by Zakharov and Shabat in \cite{ZS-JETP-1972}
where they derived conserved quantities for the nonlinear Schr\"{o}dinger equation.
Later, it was applied to the Ablowitz-Ladik hierarchy\cite{AL-JMP-1976}.
The approach was also used to obtain conserved quantities for the KdV equation (see \cite{NMPZ-book-1984}).
% but that procedure was not rigorous.
However, for the fully discrete multi-dimensionally consistent lpKdV equation \eqref{H1},
particularly, $u$ having non-zero background,
we need to reconstruct a new discrete Riccati equation by using a conformal map
which transforms the upper complex plane to the inside of unit circle.
Series solution to the Riccati equation is constructed based on the analytic and asymptotic properties of Jost solutions.

The letter is organized as follows. In Sec.\ref{sec-2} we recall some necessary results in the direct scattering problem of the lpKdV equation
given in \cite{BJ-IP-2010}.
Then in Sec.\ref{sec-3} we derive infinitely many conserved quantities for the lpKdV equation.
Finally, Sec.\ref{sec-4} is for conclusions.
Besides, we will  revisit solutions to the Riccati equation in Appendix.

\section{Preliminary}\label{sec-2}

In this section we recall some results in the IST procedure of the lpKdV equation described in \cite{BJ-IP-2010}.
These results are necessary in our approach.
Throughout the paper we suppose the spacing parameters $p>0$ and $q>0$.

The lpKdV equation \eqref{H1} has the following Lax pair
\begin{subequations}\label{Lax-H1}
\begin{align}
& \left(\begin{array}{c}
        \t\phi_1\\
        \t\phi_2
        \end{array}\right)
  = \left(\begin{array}{cc}
      -u     & u\t{u}+p^2-r^2\\
      -1     & \t{u}
      \end{array}
\right)\left(\begin{array}{c}
        \phi_1\\
        \phi_2
        \end{array}\right),\label{Lax-H1-a}\\
& \left(\begin{array}{c}
        \h\phi_1\\
        \h\phi_2
        \end{array}\right)
=\left(
\begin{array}{cc}
-u & u\h{u}+q^2-r^2\\
-1 & \h{u}
\end{array} \right)
\left(\begin{array}{c}
        \phi_1\\
        \phi_2
        \end{array}\right).\label{Lax-H1-b}
\end{align}
\end{subequations}
One can rewrite \eqref{Lax-H1-a} into a scalar discrete Schr\"odinger problem\cite{BPPS-IP-2001,LPSY-SIGMA-2008}
\begin{equation}\label{Lax-H1-Scalar-n}
  \t{\t g}- s\, \t g+(p^2+z^2)g=0,
\end{equation}
where
\begin{equation}
s=s_{n,m}=\t{\t u}-u
\label{s-u}
\end{equation}
and we have replaced the wave function $\phi_2$ with $g$ and the spectral parameter $r$ with $iz$.
It then follows from the structure \eqref{NSS} that
\begin{equation}
s \sim 2p,~~ |n|\to \infty.
\label{s-asymp}
\end{equation}
This is the boundary condition for $s$ or $u$, and we request $s$ falls in the space (cf.\cite{BJ-IP-2010,BPPS-IP-2001})
\begin{equation}\label{boundary-cond}
    P_{\mu}=\Bigl\{ s_{n,m} \,: \,\sum_{n=-\infty}^{\infty}|n^{\mu}(s_{n,m}-2p)|<\infty \Bigr\}, ~ \mu=0, 1, 2.
\end{equation}
We note that in direct scattering problem $m$ is treated as a constant or zero.

Under the boundary condition \eqref{boundary-cond}, the difference spectral problem \eqref{Lax-H1-Scalar-n}
admits Jost solutions $\varphi$ and $\psi$ with asymptotic property
\begin{subequations}\label{asymp}
\begin{align}
&\varphi(n;z)\sim (p-i z)^n,~~~(n\to -\infty), \label{asymp-1}\\
&\psi(n;z)\sim (p+i z)^{n},~~~(n\to \infty), \label{asymp-2}
\end{align}
\end{subequations}
where $\varphi(n;z)$ and $\psi(n;z)$ are analytic in the upper complex plane $\mathrm{Im}z>0$ and continuous in $\mathrm{Im} z \geq 0$.
Besides,
\begin{subequations}\label{asymp-z}
\begin{align}
&\varphi(n;z)(p-i z)^{-n} \sim 1+ O\Bigl(\frac{1}{z}\Bigr),~~~(||z||\to \infty), \label{asymp-1-z}\\
&\psi(n;z)(p+i z)^{-n}\sim 1+ O\Bigl(\frac{1}{z}\Bigr),~~~(||z|| \to \infty). \label{asymp-2-z}
\end{align}
\end{subequations}
There is a linear relation
\begin{equation}
\varphi(n;z)=a(z)\psi(n;-z)+ b(z)\psi(n;z),~~(z\in \mathbb{R}).
\end{equation}
$a(z)$ is independent of $n$ and expressed as
\begin{equation}
a(z)=\frac{1}{2iz}\mathrm{W}_n[\varphi(n;z),\psi(n;z)],
\label{a(z)}
\end{equation}
where the Casorati determinant with respect to $n$ is defined as
\begin{equation}\label{Cas}
    \mathrm{W}_n[f(n),g(n)]=(p^2+z^2)^{-n}\,\begin{vmatrix}
f(n;z) & g(n;z)\\
f(n+1;z)&g(n+1;z)
\end{vmatrix}.
\end{equation}
Thanks to the analytic property of $\varphi$ and $\psi$,
the relation \eqref{a(z)} can be extended to the upper complex plane,
i.e. $a(z)$ is analytic in $\mathrm{Im} z>0$.
For the asymptotic property,
$a(z)\sim 1$ when $||z||\to \infty$ and $a(z)$ can be expanded in the neighbourhood of $\infty$  as
\begin{equation}
a(z)=1+\sum^{\infty}_{j=1}\frac{\alpha_j}{z^j}.
\label{a(z)-exp}
\end{equation}

$a(z)$ has finite zeros $\{z=i\kappa_l\}^N_{l=1}$ and due to analytic property and uniqueness,
$a(z)$ can be expressed as (cf.\cite{BJ-IP-2010})
\[a(z)=\prod^N_{j=1}\frac{z-i\kappa_j}{z+i\kappa_j},~~~~(\kappa_j>0).\]
Therefore $a(z)$ is independent of both $n$ and $m$.

\section{Infinitely many conserved quantities}\label{sec-3}

\subsection{General scheme}\label{sec-3-1}

Making use of the relation \eqref{a(z)} and the asymptotic behavior \eqref{asymp}, we have\footnote{In the case of
the continuous KdV equation, the counterpart of \eqref{a-sim-phi} is
\[a(z)\sim \frac{e^{izx}}{2iz}(iz \varphi-\varphi_x).\]
However, the term $\varphi_x$ on the r.h.s was missed in \cite{NMPZ-book-1984}.}
\begin{equation}
a(z)\sim \frac{\varphi(n;z)}{2iz(p-iz)^n}\Bigl(p+iz-\frac{\varphi(n+1;z)}{\varphi(n;z)}\Bigr),~~ (n\to \infty),
\label{a-sim-phi}
\end{equation}
and consequently,
\begin{equation}
\ln a(z)\sim -\ln 2iz+G_n+B_n,~~(n\to \infty),
\label{a-sim-phi-ln}
\end{equation}
where
\begin{equation}
G_n=\ln \frac{\varphi(n;z)}{(p-iz)^n},~~ B_n=\ln (p+iz-\theta)
\end{equation}
and
\begin{equation}
\theta=\theta(n)=\frac{\varphi(n+1;z)}{\varphi(n;z)}.
\label{theta-1}
\end{equation}
Noting that $\lim_{n\to -\infty}B_n=\ln 2iz$, we have
\begin{equation}
\ln a(z)\sim G_n+B_n-B_{-\infty},~~(n\to \infty).
\label{a-sim-phi-ln-1}
\end{equation}
Here $B_{-\infty}$ stands for $\lim_{n\to -\infty}B_n$ and hereafter we also employ similar
notations such as $B_{\infty}$ and $G_{-\infty}$ without confusion.
Next, defining
\begin{equation}
\Gamma_n=(E_n-1)G_n =\ln \frac{\theta}{p-iz}
\end{equation}
and noting that $G_{ -\infty}=0$
due to the asymptotic property \eqref{asymp-1},
%Then, once we get $\theta$ and meanwhile if
%\begin{equation}
%\lim_{n\to -\infty} B_n =\lim_{n\to \infty} B_n,
%\end{equation}
%one will have
we find
\begin{equation}
\ln a(z) \sim \sum^{n-1}_{j=-\infty} \Gamma_j +\sum^{n-1}_{j=-\infty}( B_{j+1}-B_j),~~ (n\to \infty).
\end{equation}
Thus,  since $a(z)$ is independent of $n$ it is then recovered  via
\begin{align}
\ln a(z)=&\lim_{n\to \infty}\Bigl[\sum^{n-1}_{j=-\infty}\Gamma_j+ \sum^{n-1}_{j=-\infty}( B_{j+1}-B_j)\Bigr]\nonumber\\
=& \sum^{\infty}_{n=-\infty}\ln \frac{\theta}{p-iz}+B_{\infty}-B_{-\infty}.
\label{a-recover}
\end{align}
In principle, this will provide a set of conserved quantities $\{K_j\}$ by expanding
\begin{equation}
\ln a(z)=\ln \Bigl(1+\sum^{\infty}_{j=1}\frac{\alpha_j}{z^j}\Bigr)=\sum^{\infty}_{j=1}\frac{K_j}{z^j}.
\label{a(z)-exp-form}
\end{equation}

\subsection{Riccati equation and conformal map}\label{sec-3-2}

The known discrete Riccati equation w.r.t. $\theta$ corresponding to \eqref{Lax-H1-Scalar-n} (with $g=\varphi(n;z)$) is\cite{ZCS-JPA-2013}
\begin{equation}\label{Ric-theta}
\t{\theta}\theta- s\,\theta + \varepsilon^{2}=0,~~ \varepsilon^{2}=p^2+z^2.
\end{equation}
It admits a series solution\cite{ZCS-JPA-2013}
\begin{subequations}\label{theta}
\begin{equation}
\theta=\frac{\varepsilon^2}{s}
\Bigl(1+\sum_{j=1}^{\infty}\theta_j\varepsilon^{2j}\Bigr),
\label{theta-exp}
\end{equation}
with
\begin{equation}
\theta_{j+1}=\frac{1}{s\t s}\sum_{i=0}^{j}\t{\theta}_i\theta_{j-i},~~
j=0,1,2,\cdots,~~(\theta_0=1),\label{theta-re}
\end{equation}
\end{subequations}
in which
\begin{align}
&\theta_1=\frac{1}{s\t s},~~
\theta_2=\frac{1}{s\t s}\Bigl(\frac{1}{s\t s}+\frac{1}{\t s\,\t{\t s}}\Bigr),~~\cdots \cdots.
\end{align}

Obviously, with the above form \eqref{theta} for $\theta$, it is hard to expand $\ln \frac{\theta}{p-iz}$ into a series of $z$
with negative powers which matches the expansion \eqref{a(z)-exp-form}.
% and get the conserved quantities $\{K_j\}$.

\begin{figure}[!h]
\centering
%\subfigure[]
\includegraphics[width=3.2in,height=2.4in]{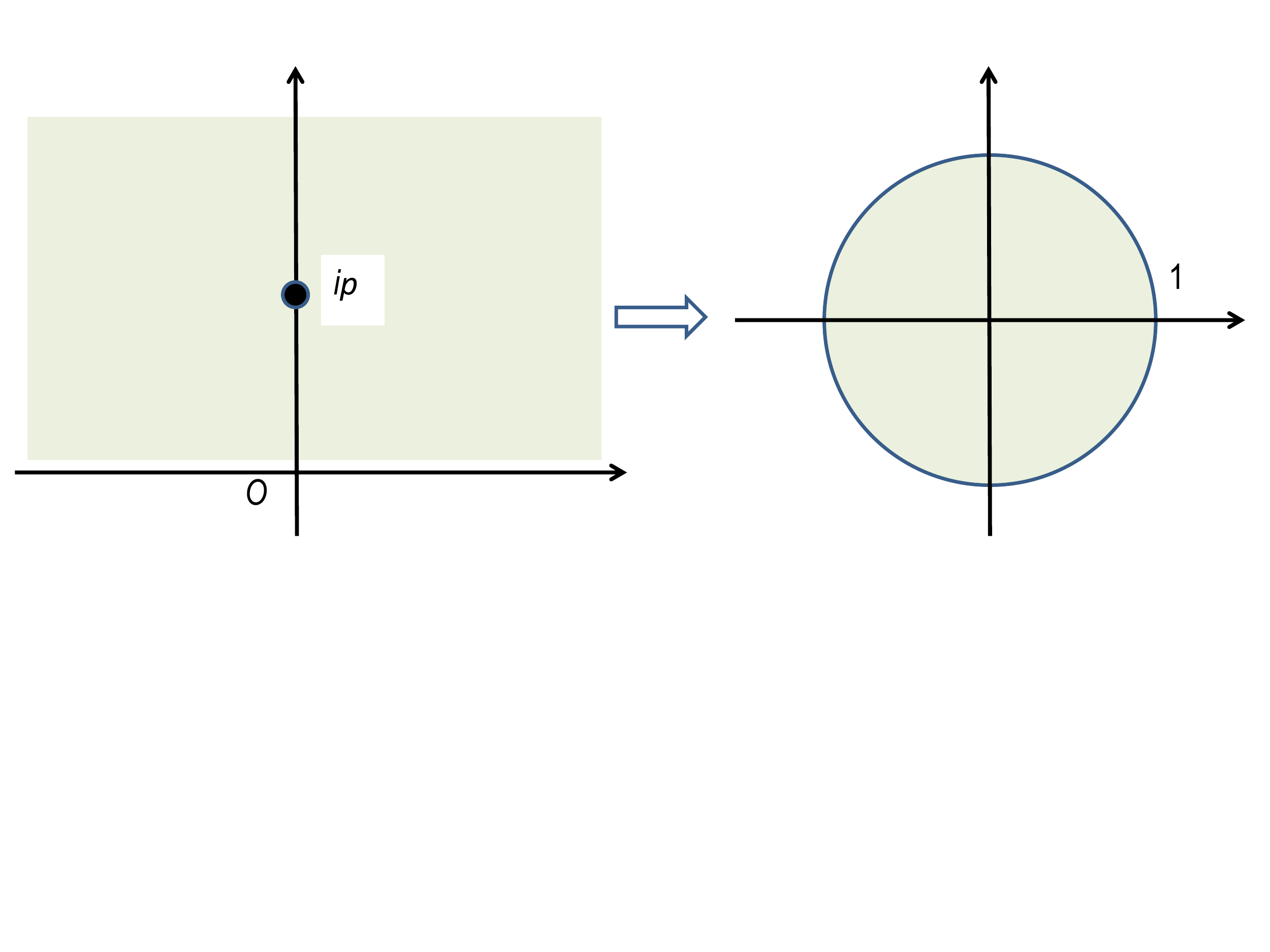}
\vskip -28mm
\caption{Conformal map \eqref{conformal}}
\label{fig:1} %% label for entire figure
\end{figure}

To over come this obstacle, we consider
\begin{equation}
\omega=\frac{\theta}{p-iz}
\label{omega-theta}
\end{equation}
under which
\begin{equation}
\Gamma_n = \ln \omega,~~~ B_n=\ln\Bigl(\frac{p+iz}{p-iz}-\omega \Bigr)+\ln (p-iz),
\label{Gamma}
\end{equation}
and the Riccati equation w.r.t. $\omega$ takes the form
\begin{equation}\label{Ric-omega}
\t{\omega}\omega- \frac{s\,\omega}{p-iz}+\frac{p+iz}{p-iz}=0.
\end{equation}
To solve it, we introduce  a conformal map
\begin{equation}
\lambda=\frac{p+iz}{p-iz}
\label{conformal}
\end{equation}
to transform the upper $z$-plane to the inside of unit circle on the $\lambda$-plane, as depicted in Fig.\ref{fig:1}.
\vskip -4mm
Under this conformal map and noting that
\begin{equation}
z=ip\frac{1-\lambda}{1+\lambda},
\label{z-lambda}
\end{equation}
\eqref{Gamma} reads
\begin{equation}
\Gamma_n = \ln \omega,~~~ B_n=\ln (\lambda-\omega  )+\ln\frac{2p}{1+\lambda},
\label{Gamma-lamb}
\end{equation}
\eqref{a-recover} reads
\begin{align}
\ln a\Bigl(ip\frac{1-\lambda}{1+\lambda}\Bigr)=\sum^{\infty}_{n=-\infty}\ln \omega+B_{\infty}-B_{-\infty},
\label{a-recover-2}
\end{align}
and the Riccati equation \eqref{Ric-omega} is written as
\begin{equation}\label{Ric-omega-lamb}
\omega(\t{\omega}- \t x)-\lambda \t x \omega+\lambda=0,
\end{equation}
where
\begin{equation}
x=\dt s /2p= (\t u-\dt u)/2p,
\label{x}
\end{equation}
which tends to 1 as $|n|\to \infty$.
\eqref{Ric-omega-lamb} has the following solution,
\begin{subequations}\label{omega}
\begin{equation}
\omega=x \Bigl(1+\sum_{j=1}^{\infty}\omega_j \lambda^j\Bigr),~~~(||\lambda||<1)
\end{equation}
with
\begin{align}
%& \omega_1=1-\frac{1}{x_{n,m}x_{n-1,m}}\\
& \omega_1 = 1-\frac{1}{x\,\dt x}, \\
& \omega_2 = \Bigl(1-\frac{1}{\dt x\,\dt{\dt x}}\Bigr)\,\frac{1}{x\,\dt x}, \\
& \omega_{j+1}=\dt \omega_j - \sum_{k=1}^{j}\dt\omega_k \, {\omega}_{j+1-k},~~ j=2,3,\cdots.\label{omega-re}
\end{align}
\end{subequations}
Noting that the fact $x\sim 1$ as $|n|\to \infty$, we immediately have the following.
\begin{prop}\label{P:1}
All the $\{\omega_j\}$ defined in \eqref{omega} are asymptotic to zero as $|n|\to \infty$.
\end{prop}

\subsection{Infinitely many conserved quantities}\label{sec-3-3}

From now on everything is considered on $\lambda$-plane.
Based on proposition \ref{P:1}, for the term $B_n$ in \eqref{Gamma-lamb} we find
$B_{\infty}=B_{-\infty}$,
%\[\lim_{n\to -\infty}\ln(\lambda-\omega)= \lim_{n\to \infty}\ln(\lambda-\omega),\]
%which means $\sum^{\infty}_{n=-\infty}\Gamma_n = \sum^{\infty}_{n=-\infty} \ln \omega$.
%Now, combining \eqref{a-recover}, \eqref{Gamma-lamb}, \eqref{omega} and \eqref{conformal} or \eqref{z-lambda}, together with the above proposition, we reach
from which and \eqref{a-recover-2} it follows that
\begin{equation}
\ln a\Bigl(ip\frac{1-\lambda}{1+\lambda}\Bigr)=\sum^{\infty}_{n=-\infty}\ln \omega.
\label{a-ome-ln}
\end{equation}
Theoretically, the l.h.s. has the expansion (at the origin point of $\lambda$-plane)
\begin{equation}
\ln a\Bigl(ip\frac{1-\lambda}{1+\lambda}\Bigr)=H_0 + \sum^{\infty}_{j=1}H_j\, \lambda^j
\label{a(z)-exp-lam}
\end{equation}
due to the property that $a\bigl(ip\frac{1-\lambda}{1+\lambda}\bigr)$ is analytic inside the unit circle,
while the r.h.s. of \eqref{a-ome-ln} can be explicitly expressed as
\begin{align}
\sum^{\infty}_{n=-\infty}\ln \omega
 = & \sum^{\infty}_{n=-\infty}\ln x
      + \sum^{\infty}_{n=-\infty}\ln \Bigl(1+\sum_{j=1}^{\infty}\omega_j \lambda^j\Bigr)\nonumber\\
 = & \sum^{\infty}_{n=-\infty}\ln x + \sum_{j=1}^{\infty}\left(\, \sum^{\infty}_{n=-\infty} h_j(\mathbf{\bom})\right)  \lambda^j.
 \label{omega-ln}
\end{align}
Here, $\{h_j(\mathbf{\bom})\}$, we call them $h$-polynomials for short,
are the polynomials defined in the following way (cf.\cite{ZCS-JPA-2013})
%\begin{subequations}\label{exp-t}
\begin{equation}
\ln\biggl(1+\sum_{i=1}^{\infty}\omega_i \lambda^i
\biggr)=\sum_{j=1}^{\infty}h_j(\bom)\lambda^{j}, \label{hom}
\end{equation}
where $\bom=(\omega_1,\omega_2,\cdots)$ and the first three of
 $\{h_j(\bom)\}$ are
%\begin{subequations}\label{ht1-4}
\begin{align}
h_1(\bom)=\omega_1,~~ h_2(\bom)=-\frac{1}{2}\omega_1^2+\omega_2, ~~ h_3(\bom)=\frac{1}{3}\omega_1^3-\omega_1\omega_2+\omega_3.
%\\
%& h_4(\bft)=-\frac{1}{4}t_1^4+t_1^2t_2-t_1t_3-\frac{1}{2}t_2^2+t_4.
\end{align}
%\end{subequations}
% $\{h_j(\mathbf{2})\}$ are defined as above with $\mathbf{2}=(2,2,\cdots)$.
A general description for such polynomials is given in Appendix \ref{A:1}.

Thus, it is clear that the infinitely many conserved quantities are the following.
\begin{prop}\label{P:2}
The infinitely many conserved densities of the lpKdV equation \eqref{H1} are
\begin{equation}
H_0= \sum^{\infty}_{n=-\infty}\ln x ,~~ H_j=  \sum^{\infty}_{n=-\infty} h_j(\mathbf{\bom}),
\label{cqs}
\end{equation}
and the corresponding conserved densities are
\begin{equation}
F_0=\ln x ,~~ F_j= h_j(\mathbf{\bom}),
\label{cds}
\end{equation}
where $x= (\t u-\dt u)/2p$ and the $h$-polynomials are defined in Appendix \ref{A:1}.
%We note that the constant terms $\ln(-2p)$ and $h_j(\mathbf{2})$ can be removed.
\end{prop}

For these conserved quantities or conserved densities, we have further results.
\begin{prop}\label{P:3}
Under the criteria given in \cite{MWX-TMP-2011}, the conserved densities \eqref{cds} are non-trivial and not equivalent to each other.
\end{prop}
This can be proved as done in \cite{ZCS-JPA-2013} by making use of $h$-polynomials to calculate orders of conserved densities.

\vskip 8pt
At the end of the section we note that
it is not necessary to compare the conserved densities \eqref{cds} with
\begin{equation}
F'_0=-\ln (u-\t{\t u}) ,~~ F'_j= h_j(\mathbf{\bthe}),~~j=1,2,\cdots
\label{cds-old}
\end{equation}
which were derived in \cite{ZCS-JPA-2013} based on the series solution $\theta$ given in \eqref{theta}. Here $\bthe=(\theta_1,\theta_2,\cdots)$.
In fact, on one hand, $F_j$ are functions of $\{u_{n+1-s}\}$
while $F'_j$ are functions of $\{u_{n+s}\}$ where $s=0,1,2,\cdots$.
They can not be unified.
On the other hand, any series solutions to the Riccati equation \eqref{Ric-theta} should agree with known analytic properties.
In this sense, the expansion \eqref{theta-exp} is \textit{wild}.

%We will leave a section in Appendix to discuss series solutions to  the Riccati equation \eqref{Ric-omega-lamb}.

A further discussion on the series solutions to the Riccati equation \eqref{Ric-omega-lamb}
is given in Appendix \ref{A:2}.

\section{Conclusions}\label{sec-4}

We have derived infinitely many conserved quantities for the lpKdV equation \eqref{H1}.
%Another set of conserved quantities with respect to $(m,p)$ can be obtained by replacing $(n,p,\t{}\,\,)$ with $(m,q,\h{}\,\,)$ in proposition \ref{P:2}.
Different from those known ones, these conserved quantities are convergent thanks to the zero asymptotic behaviors of conserved densities.
They are also nontrivial and not equivalent to each other.
The derivation is based on analytic and asymptotic properties
of Jost solutions of the discrete Schr\"{o}dinger spectral problem \eqref{Lax-H1-Scalar-n}.
The conformal map \eqref{conformal} is employed to convert the upper $z$-plane to the inside unite circle of $\lambda$-plane.
Such a spectral transformation enables us to have a new series solution of the Riccati equation with respect to $\lambda$
and the solution agrees with the analytic and asymptotic properties
of Jost solution $\varphi(n;z)$.
This new solution leads to desired infinitely many conserved quantities, which presented in Proposition \ref{P:2}.
Another set of conserved quantities with respect to $(m,q)$ can be obtained by replacing $(n,p,\t{}\,\,)$ with $(m,q,\h{}\,\,)$.

It has been seen that the derivation of conserved quantities or conservation laws for discrete systems is more complicated than continuous case.
One reason is due to the obstacle of non-zero background of soliton solutions in discrete case.
It is specially important to construct conserved densities which are  asymptotic to zero when solutions have non-zero background.
Another point is Riccati equation and its solutions. For the lpKdV equation, we have three versions of its Riccati equation, which are
\eqref{Ric-theta}, \eqref{Ric-omega} and \eqref{Ric-omega-lamb}. Even for \eqref{Ric-omega-lamb} that we finally used in the paper,
it has two different series solutions (see Appendix \ref{A:2}) but only one agrees with the analytic and asymptotic properties of Jost solution $\varphi$.
We hope by the discussion of the letter we can emphasize the importance of asymptotic property and analytic property of discrete wave functions
in the research of discrete systems.

It is possible to examine asymptotic behavior  of solutions of other equations in the ABS list
which are solved via IST\cite{Butler-Non-2012},
and make use of analytic and asymptotic properties of Jost solutions to derive reasonable conserved quantities.
It is also possible to re-examine Lax pair approach in \cite{ZCS-JPA-2013} and construct reasonable Riccati equations
and their series solutions for those multi-dimensionally consistent lattice equations
(see the list in \cite{BHQK-FCM-2013}).
%These will be considered elsewhere.

\vskip 15pt
\subsection*{Acknowledgments}
The authors  are very grateful to Prof. Dengyuan Chen for his enthusiastic discussion.
This project is supported by the NSF of China (Nos. 11435005, 11175092, 11205092, 11371241) and SRF of
the DPHE of China (No. 20113108110002).

\vskip 20pt

\appendix
\section{$h$-polynomials\cite{ZCS-JPA-2013}}\label{A:1}

The following expansion holds,
\begin{subequations}\label{exp-t}
\begin{equation}
\ln\biggl(1+\sum_{i=1}^{\infty}t_i k^i
\biggr)=\sum_{j=1}^{\infty}h_j(\bft)k^{j}, \label{ht}
\end{equation}
where
\begin{equation}
h_j(\bft)=\sum_{||\balpha||=j}(-1)^{|\balpha|-1}(|\balpha|-1)!\frac{\bft^{\balpha}}{\balpha
!}, \label{htj}
\end{equation}
and
\begin{align}
& \mathbf{t}=(t_1,t_2,\cdots),~~\balpha=(\alpha_1,\alpha_2,\cdots),~~\alpha_i\in\{0,1,2,\cdots\}, \\
& \bft^{\balpha}=\prod_{i=1}^{\infty}t_i^{\alpha_i},~~
 {\balpha}!=\prod_{i=1}^{\infty}(\alpha_i !),~~
|\balpha|=\sum_{i=1}^{\infty}\alpha_i,~~
||\balpha||=\sum^{\infty}_{i=1} i\alpha_i.
\end{align}
\end{subequations}
The first few of $\{h_j(\bft)\}$ are
\begin{subequations}\label{ht1-4}
\begin{align}
& h_1(\bft)=t_1,\\
& h_2(\bft)=-\frac{1}{2}t_1^2+t_2, \\
& h_3(\bft)=\frac{1}{3}t_1^3-t_1t_2+t_3,\\
& h_4(\bft)=-\frac{1}{4}t_1^4+t_1^2t_2-t_1t_3-\frac{1}{2}t_2^2+t_4.
\end{align}
\end{subequations}

\section{Proper solution to the Riccati equation \eqref{Ric-omega-lamb}}\label{A:2}

In this section, we investigate how a proper series solution of the Riccati equation \eqref{Ric-omega-lamb}
is constructed based on the analytic and asymptotic properties of the Jost solution $\varphi(n;z)$.

For the Riccati equation \eqref{Ric-omega-lamb}, in addition to the series solution \eqref{omega}, it has a second solution:
\begin{subequations}\label{omega-2nd}
\begin{equation}
\omega=\sum_{j=1}^{\infty}\omega'_j \lambda^j,
\end{equation}
with
\begin{align}
%& \omega_1=1-\frac{1}{x_{n,m}x_{n-1,m}}\\
& \omega'_1 = \frac{1}{\t x},~~
%& \omega_2 = \Bigl(1-\frac{1}{\dt x\,\dt{\dt x}}\Bigr)\,\frac{1}{x\,\dt x}, \\
 \omega'_{j+1}=- \omega'_j + \frac{1}{\t x} \sum_{k=1}^{j}\t\omega'_k \, {\omega}'_{j+1-k},~~ j=1,2,\cdots.\label{omega-re-2}
\end{align}
\end{subequations}
This solution has asymptotic property $\omega \sim \lambda$ as $|n|\to \infty$,
but the expansion is `wild' because it
does not agree with the analytic and asymptotic properties of $\varphi(n;z)$.

Let us explain this in the following.
In the light of asymptotic behavior \eqref{asymp-1-z} and noting that \eqref{omega-theta}, i.e.
$\omega=\frac{\varphi(n+1;z)}{\varphi(n;z)}\frac{1}{p-iz}$,
there is
\begin{equation}
\omega \sim 1+O\Bigl( \frac{1}{z}\Bigr),~~~ ||z||\to \infty,
\label{omega-asym}
\end{equation}
which means  $\omega$ can be expanded at $\infty$ as
\begin{equation}
\omega = 1+\sum^{\infty}_{j=1}\frac{\omega^{(j)}}{z^j}.
\label{omega-z}
\end{equation}
Meanwhile, the conformal map \eqref{conformal} indicates
\begin{equation}
\frac{1}{z}=\frac{1}{ip}(1+ 2\sum^{\infty}_{j=1}\lambda^j),
\end{equation}
and consequently,
\begin{equation}
\frac{1}{z^j}=\frac{1}{(ip)^j}(1+O(\lambda)).
\end{equation}
This means, by substituting the above into \eqref{omega-z}, $\omega$ should have the following expansion in terms of $\lambda$:
\begin{equation}
\omega = \omega_0+\sum^{\infty}_{j=1} \omega_j \lambda^j
\label{omega-lambda}
\end{equation}
and the term $\omega_0$ is
\begin{equation}
\omega_0=1+\sum^{\infty}_{j=1}\frac{\omega^{(j)}}{(ip)^j}.
\end{equation}
This is nothing but $\omega(n;z=ip)$ in light of the expression \eqref{omega-z}.
To calculate $\omega(n;ip)$ we take $z=ip$ in the spectral problem \eqref{Lax-H1-Scalar-n} with $g=\varphi(n;z)$,
we have
\[\varphi(n+2;ip)-2p \t x \varphi(n+1;ip)=0,\]
which gives
\begin{equation}
\omega(n;ip)=\frac{\varphi(n+1;ip)}{\varphi(n;ip)}\frac{1}{2p}=x.
\end{equation}
Such a value, together with the expression \eqref{omega-lambda},
coincides with the expansion \eqref{omega}.

Thus, it is clear that the expansion \eqref{omega} is actually well-based on the analytic and asymptotic properties
of the wave function $\varphi(n,z)$ while the expansion \eqref{omega-2nd} not.
In other words, the analytic and asymptotic properties of wave functions play important roles in the procedure to find conserved quantities
for discrete integrable systems.

%\end{CJK*}

\end{document}